\documentclass[a4paper,pt,twoside]{cpc-hepnp}

\usepackage{multicol}
\usepackage{graphicx}
\usepackage{booktabs}
\usepackage{amssymb,bm,mathrsfs,bbm,amscd}
\usepackage[tbtags]{amsmath}
\usepackage{lastpage}
\usepackage{CJK}
\usepackage{hyperref}
\usepackage{url}
\usepackage{float}
\usepackage{prettyref}
\usepackage{natbib}
\newrefformat{fig}{Fig. \ref{#1}}
\newrefformat{tab}{Tab. \ref{#1}}
\newrefformat{equ}{formula \ref{#1}}

\begin{document}
\begin{CJK*}{GBK}{kai}
\bibliographystyle{plain}

%\fancyhead[co]{\footnotesize XU Ming~ et al: Comparison of Thermal Neutron Detection Efficiency of $^{6}$Li Scintillation Glass and $^{3}$He
%Gas Proportional Tube}

\footnotetext[0]{Submitted to Chinese Physics C}

\title{Comparison of Thermal Neutron Detection Efficiency of $^{6}$Li Scintillation Glass and $^{3}$He
Gas Proportional Tube}

\author{%
      XU Ming$^{1,2}$%(徐明)%
\quad TANG Zhi-Cheng$^{1;1)}$%(唐志成)
\email{tangzhch@mail.ihep.ac.cn}%
\quad CHEN Guo-Ming$^{1}$%(陈国明)%
\quad TAO Jun-Quan$^{1}$%(陶军全)%
 }

\maketitle

\address{%
 $^{1}$Institute of High Energy Physics, Chinese Academy of Sciences, Beijing 100049, China\\
 $^{2}$University of Chinese Academy of Sciences, Beijing 100049, China
}

\begin{abstract}
We report on a comparison study of the $^{3}$He gas proportional
tube and the $^{6}$Li incorporated scintillation glasses on thermal
neutron detection efficiency. Both $^{3}$He and $^{6}$Li are used
commonly for thermal neutron detection because of their high neutron
capture absorption coefficient. By using a neutron source $^{252}$Cf
and a paraffin moderator in an alignment system, we can get a small
beam of thermal neutrons. A flash ADC is used to measure the thermal
neutron spectrum of each detector, and the detected number of events is
determined from the spectrum, then we can calculate the detection
efficiency of different detectors. Meanwhile, the experiment have
been modeled with GEANT4 to validate the results against the Monte
Carlo simulation.
\end{abstract}



\begin{multicols}{2}
\section{Introduction}
Space activities have been increased considerably in the past
decades\cite{space_review}. Electron/hadron discrimination power
is an important factor for space experiments designed to study
cosmic ray at high energy. Electromagnetic calorimeters have good
intrinsic electron/hadron rejection power based on the shower
topology, and this property requires the calorimeters with high
granularity read out and sufficient depth so that the showers can
fully develop. Unfortunately, in space experiments, due to power
consumption limitations and weight constraints, the electron/hadron
discrimination capability is reduced consequently. The
number of secondary neutrons in the hadronic shower is much larger than that in
the electromagnetic shower, therefore to place a neutron detector downstream
of the calorimeter, aimed to exploit the neutron components of the
shower is a feasible solution to enhance the electron/hadron
discrimination capability. Neutrons produced in
the shower are with typical kinetic energies of a few
MeV\cite{shower_review}. Generally, there are two kinds of method
for the neutron detection.
One is the fast neutron detection with elastic scattering, which neutrons lose their kinetic energy at their initiated energies.
The other is thermal neutron detection, neutrons captured by nucleus after they have lost practically all
their kinetic energy through the thermalization process.
Fast neutron detection requires low background environment, because
the ionization signal generated from elastic scattering is small,
and considering the neutron
detector is to be put downstream of the electromagnetic calorimeter,
a large number of charged particle can be produced in the shower as
the background to neutron detector, it is difficult to pick up the neutron signal from such a background. The thermal neutron detection is based on nuclear reaction, the signal generated from neutron absorbtion is much larger than the ionization signal of the charged particles, it is
a more appropriate solution.

Many type of thermal neutron detector have been developed, like gas
filled tube, liquid organic scintillator and solid state
scintillator\cite{neutron_review}\cite{neutron_review_book}. Thermal
neutron detection is based on nuclear reactions. Because of their
high neutron capture absorption coefficient, both isotope of
$^{3}$He and $^{6}$Li are used commonly for detecting thermal
neutron. $^{3}$He gas proportional tube is a typical thermal neutron
detector, the reaction when a neutron is absorbed by $^{3}$He is:
\begin{eqnarray}
\mathrm{n} + ^{3}\mathrm{He} \rightarrow ^{3}\mathrm{H} + \mathrm{p} + \mathrm{0.764\ MeV.}
\end{eqnarray}
The reaction produces a triton and a proton, which are emitted in
opposite directions and cause ionizations in the tube. Pulse from the
proportional tube is fed into a preamplifier, which enlarging the
signal for the following up device to record. Lithium glass is a
scintillation type detector, for neutron detection, often enriched
with the isotope $^{6}$Li, the nuclear reaction of $^{6}$Li by
capturing a thermal neutron, release a triton and an alpha particle:
\begin{eqnarray}
\mathrm{n} + ^{6}\mathrm{Li} \rightarrow ^{3}\mathrm{H} + ^{4}\mathrm{He} + \mathrm{4.75\ MeV.}
\end{eqnarray}
Lithium glass thus emit light in response to excitation energy
received from ionizing radiation, and the output is usually
collected by a photomultiplier.

GEANT4 is a widely used toolkit for the simulation of the passage of
particles through matter\cite{geant4}. Its fields of application
include high energy, nuclear and accelerator physics, as well as
studies in medical and space science. As such, it can handle many
physically possible process, including decay, nuclear interaction
and particle production. GEANT4 includes facilities for handing
geometry, tracking, detector response, run management, visualization
and user interface, the detector response is recording when a
particle passes through the volume of detector, and use a suitable
physics list approximating how a real detector would respond. For
many physics processes, this means less time need be spent on the
low level details, and user can start immediately on the more
important aspects of the simulation. The geometry of the efficiency
measurement experiment have been accurately modeled with GEANT4,
including the physical layout, detectors, and the materials.
\section{Experiment and Simulation Setup}
\subsection{Experiment}
The $^{3}$He tube used in this experiment is made by GE. It has a 1
inch inner diameter and 1 meter length, with filled pressure 10 atm
including 80\% $^{3}$He gas and 20\% Isobutane gas, and 304
stainless steel as the outer shell. A Cadmium coat is covered from
both ends of the tube as the thermal neutron absorber, only a length
of 8 cm in the middle of the tube left, exposed as the sensitive
area.

The Lithium glasses are of type ST-602, made by CNNC Beijing Nuclear
Instrument Factory. There are three round pieces with thickness 1 mm, 2 mm
and 3 mm, the diameter of these glasses are 2 cm.

As the two kinds of detector have different geometries, the best
solution for detection efficiency comparison is to use a small area
of thermal neutron beam. In this experiment, by assembling a neutron
source $^{252}$Cf and a paraffin moderator in an alignment system
made of paraffin barrel, which is covered with thermal neutron
absorber, we can get the thermal neutron flow from this system. The
radius of the paraffin barrel is 29 cm, with the inner 24 cm part
made by paraffin and the outer 5 cm part made by Boron-containing
paraffin, the height of the barrel is 60 cm. In the middle of the
barrel, a collimation hole with dimension of 2 cm radius and 20 cm height
is scooped, where the $^{252}$Cf source and a 5 cm long paraffin moderator are placed. There are 20 layers of 1 mm thickness
Boron-containing plastic, each layer with 1 cm radius hole in the center ,
placed on the top of the paraffin barrel as the thermal neutron
absorber and alignment.
\begin{figure}[H]
\centering
\includegraphics[width=7.5cm]{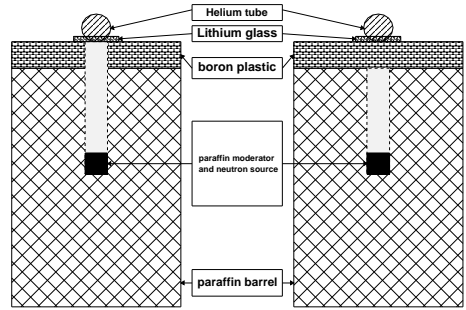}
\caption{\label{fig:1} Schematic diagram of the experiment setup}
\end{figure}
$^{3}$He tube and Lithium glasses are put on the top center of the
paraffin barrel separately for each measurement. In the case of
$^{3}$He, a pair of preamplifiers are connected to both ends of the
detector, because the magnification factor of the detector itself is
very weak. After the preamplifier, a fan in/out module is used to
add the signal from both end of the detector, and the output is
copied to two channels, one for trigger, another for spectrum
measurement. For Lithium glasses measurement, the setup is almost the
same as the $^{3}$He tube, only except for without the preamplifier,
because a photon multiplier tube is applied as the Lithium glass
readout, and this kind of detector have sufficient magnification
factor. Finally, a Flash ADC is used to record the pulse shape of
each event.

For each kind of detector, two sets of measurement are implemented for 120 minutes. Firstly, the apertural Boron-containing plastic is
placed at the center of the barrel, with positional relationship
shown in the left plot of \prettyref{fig:1}., both signal from the collimation
hole and the background are collected. Then these Boron-containing
plastic is replaced with another set of complete plastic plates of the same thickness,
as showed in the right plot of \prettyref{fig:1}., and fully covers the top of the paraffin barrel, the positional relationship of the detectors and the paraffin barrel is unchanged. In this case, only the background is collected. The only difference between these
two measurements is the thermal neutron flux from the central hole. Therefore, the difference of these two measurements represents the contribution from the center collimation hole of the plastic plates, and it makes a thermal neutron beam-like source with this kind of setup.
\begin{figure}[H]
\includegraphics[width=7.0cm]{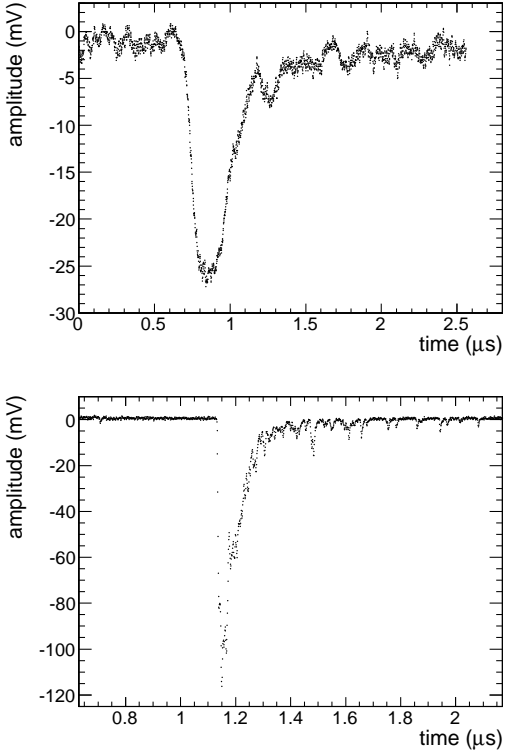}
\figcaption{\label{fig2} Upper one and lower one respectively represent
$^{3}$He tube pulse shape and 1 mm $^{6}$Li glass pulse shape recorded by
Flash ADC }
\end{figure}
\subsection{Monte Carlo Simulation}
The experiment setup have been accurately modeled with GEANT4 to
validate the results with the Monte Carlo simulation. The same
experimental geometry and materials used in the measurement was modeled within GEANT4. A complete list of materials
used is given in \prettyref{tab:1}..
\begin{table}[H]
\caption{\label{tab:1}  A description of the composition of
materials in the simulation}
\scriptsize
\begin{tabular*}{80mm}{c@{\extracolsep{\fill}}cccc}
\toprule
Material & $\rho$ (g/cm$^{3}$) & Element& A(g/mol) & Fraction(mass)\\
\hline
Paraffin & 0.93 & H & 1.01 & 0.15\\
         &      & C & 12.01 & 0.85\\
\hline
Boron & 1.03 & H & 1.01 & 0.08\\
  Pastic              &      & C & 12.01 & 0.87\\
              &      & B & 10.81 & 0.05\\
\hline
Lithium & 2.8 & O & 16.00 & 0.44\\
Glass              &      & Na & 22.99 & 0.09\\
              &      & Si & 28.09 & 0.32\\
              &      & Ca & 40.08 & 0.11\\
              &      & Li & 6.90 & 0.35\\
              &      & Li & 7.98 & 0.05\\
\hline
$^{3}$He Gas  & 0.0089 & He & 3.00 & 1\\
\hline
Isobutane    & 0.005 & H & 1.01 & 0.17\\
  Gas              &      & C & 12.01 & 0.83\\
\hline
Steel         & 7.5 & Fe & 55.85 & 1\\
\bottomrule
\end{tabular*}
\end{table}
Experiment comparison with GEANT4 simulations are done at the
level of the energy deposition in Lithium glasses and $^{3}$He tube,
without modeling of readout chain response in the simulation.
GEANT4 provides a user-customizable physics list that may be changed
to suit the requirement of the simulation. The simulated samples
have been produced using physics list suitable for the
case\cite{QGSP}: QGSP\_BIC\_HP, with Binary Cascade model and
Quark-Gluon String Precompound model to generate the final state for
hadron inelastic scattering respectively below 10 GeV, with addition
to use the data driven high precision neutron package to transport
neutrons below 20 MeV down to thermal energies. With the same setup as in the measurement, a typical size of 1 million events with isotropic emission of
4$\pi$ solid angle is generated each time. The emission spectrum of the neutron source as showed in \prettyref{tab:2}.
\begin{table}[H]
\caption{\label{tab:2}  MC neutron emission spectrum}
\scriptsize
\begin{tabular*}{80mm}{c@{\extracolsep{\fill}}cc}
\toprule Energy (MeV) & Relative Intensity\\
\hline
0.1 \hphantom{00} & \hphantom{0} 0.06\\
0.5 \hphantom{00} & \hphantom{0} 0.45\\
1.0 \hphantom{00} & \hphantom{0} 0.90\\
2.0 \hphantom{00} & \hphantom{0} 1.0\\
3.0 \hphantom{00} & \hphantom{0} 0.84\\
4.0 \hphantom{00} & \hphantom{0} 0.63\\
5.0 \hphantom{00} & \hphantom{0} 0.41\\
6.0 \hphantom{00} & \hphantom{0} 0.24\\
7.0 \hphantom{00} & \hphantom{0} 0.13\\
8.0 \hphantom{00} & \hphantom{0} 0.09\\
9.0 \hphantom{00} & \hphantom{0} 0.96\\
10.0 \hphantom{00} & \hphantom{0} 0.92\\
\bottomrule
\end{tabular*}
\end{table}
\section{Results and discussion}
\subsection{Event number counting}
\begin{figure}[H]
\includegraphics[width=7.5cm,height=6.9cm]{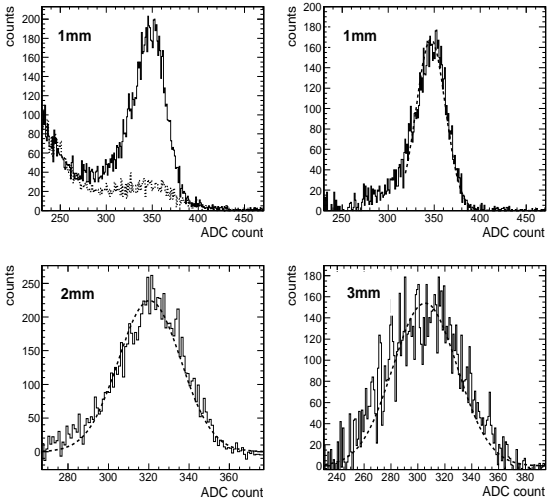}
\caption{\label{fig:4} Neutron spectrum in experiment obtained by using
Lithium glasses. Solid and dashed line in the upper left plot are
respectively the spectrum of 1 mm Lithium glass in the left case and
right case of \prettyref{fig:1}. The rest three plots are the substraction
spectrums of 1 mm, 2 mm and 3 mm Lithium glass, the dashed lines are
the Gaussian fitting functions, the number of events under the fitted Gaussian function represents the number of thermal neutron events from the collimation hole.}
\end{figure}

By integrating the pulse recorded from the flash ADC, we can get the
spectrum. With the spectrum calculated from the left case minus the
right case of \prettyref{fig:1}., we can get the signal from only the
collimation hole. In the $^{3}$He tube case, because neutron pulse shape is much wider
than the background pulse shape, only signal with pulse width greater than 200 ns is integrated for the
spectrum, the detected number of events can be calculated by adding the counts of the substraction
spectrum. Lithium glasses suffer from the photomultiplier noise
and other radiation, so using Gaussian function
to fit the substraction spectrums is a solution, as showed in \prettyref{fig:4}., and
the number of events under the fitted Gaussian function is the detected number of events from the collimation hole.
\begin{figure}[H]
\includegraphics[width=7.5cm,height=6.95cm]{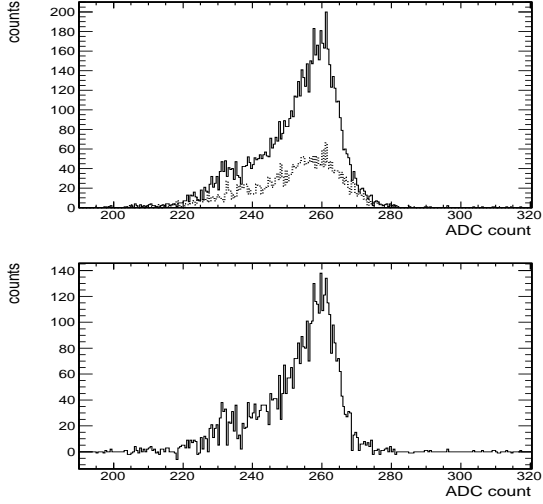}
\figcaption{\label{fig4} Neutron spectrum in experiment obtained by using
$^{3}$He tube. Solid and dashed line are respectively the spectrum
of $^{3}$He tube in the left case and right case of \prettyref{fig:1}., the
lower plot is the solid line minus the dashed of the upper plot,
which means the spectrum of thermal neutron only from the
collimation hole.}
\end{figure}
The efficiency of the detector is defined as the ratio of the number
of detected neutron induced interactions divided by the number of
incident neutrons. In this paper, We determine the number of detected thermal
neutron from the measured spectrums, and calculate the number of incident
thermal neutrons  by fitting of detection efficiency
versus different detector thickness of Lithium glasses with
experiment data. The half life of $^{252}$Cf is about 2.6 years, so we can assume radioactive source strength is unchanged in all measurements. For the Lithium glasses of
different thickness, there is a relationship between the detection
efficiency and detector thickness\cite{McGregor2003272}:
\begin{equation}
\eta(d) = 1 - (1 - \eta_{1})^{d},
\end{equation}
\begin{equation}
N_{det}(d) = N_{0}[1- (1 - \eta_{1})^{d}],
\label{equ:4}
\end{equation}
\begin{math}
\eta(d)\
\end{math}
stands for the efficiency of Lithium glass with thickness
\begin{math}
\mathit{d},\ N_{det} \mathrm{\ and\ } N_{0}
\end{math}
regard to neutron induced events and
incident neutron events respectively. For the absolute efficiency,
\begin{math}
N_{0}
\end{math}
can be determined by the fitting of the function with the
experiment data of different thickness Lithium glass scintillator
\subsection{Systematic Error Estimation}
The diameters of the collimator hole and the $^{3}$He tube are 2 cm
and 2.5 cm respectively, it means that the geometry relationship of
the collimator hole and the barrel is the most sensitive factor to the
systematic error. The $^{3}$He tube is measured several rounds, and using the standard deviation
of the measurements as the event number uncertainty:
\begin{equation}
\Delta{N_{^{3}He}} = \sqrt{\frac{\sum^{n}_{i}(N_{i}-\overline{N})^{2}}{n}}.
\end{equation}

The area of lithium glasses are much larger than the
collimator hole, the spatial relationship will not introduce much
systematic error. The detected thermal neutron number of lithium
glasses are determined by the fitting of the measured spectrums as showed in \prettyref{fig:4}., that means the
uncertainty of fitting is the most sensitive factor to the
systematic error, the fitting error can be treated as
the event number uncertainty $\Delta{N_{^{6}Li}}$.
The total incident number of events $N_{0}$ and event number
uncertainty ${\Delta{N_{0}}}$ can be determined by the fitting of
\prettyref{equ:4}. with the experiment data of Lithium glasses. The
detection efficiency is the ratio of detected number of events
divided by the number of incident number of events:
\begin{equation}
\epsilon_{det} = \frac{N_{det}}{N_{0}},
\label{equ:6}
\end{equation}
and the uncertainty on the efficiency $\epsilon_{det}$
is\footnote{since ${N_{det}}$ is a subset of ${N_{0}}$, the correlation between ${N_{det}}$ and ${N_{0}}$ is ${N_{det}}$, the covariance matrix of (${N_{det}}$,${N_{0}}$) will be in the form of
cov(${N_{det}}$,${N_{det}}$) = $({\Delta{N_{det}}})^{2}$, cov(${N_{0}}$,${N_{0}}$) = $({\Delta{N_{0}}})^{2}$, cov(${N_{det}}$,${N_{0}}$) = cov(${N_{0}}$,${N_{det}}$) = $({\Delta{N_{det}}})^{2}$, with the derivative $\frac{\partial{\epsilon_{det}}}{\partial{N_{0}}}$ = $\frac{-\epsilon_{det}}{N_{0}}$, $\frac{\partial{\epsilon_{det}}}{\partial{N_{det}}}$ = $\frac{1}{N_{0}}$, and $(\Delta{\epsilon_{det}})^{2}$ = $\mathrm{cov}^{2}(\epsilon_{det})$ = $[(1-2\epsilon_{det})(\Delta{N_{det}}^{2})+\epsilon_{det}^{2}(\Delta{N_{0}}^{2})]/N_{0}^{2}$\cite{error_book}.}:
\begin{equation}
\Delta{\epsilon_{det}}=\frac{1}{N_{0}}\sqrt{(1-2\epsilon_{det})(\Delta{N_{det}}^{2})+\epsilon_{det}^{2}(\Delta{N_{0}}^{2})}.
\end{equation}
The systematic error of the detection efficiency $\Delta{\epsilon_{det}}$ is showed in \prettyref{tab:3}.
\begin{figure}[H]
\includegraphics[width=8.5cm]{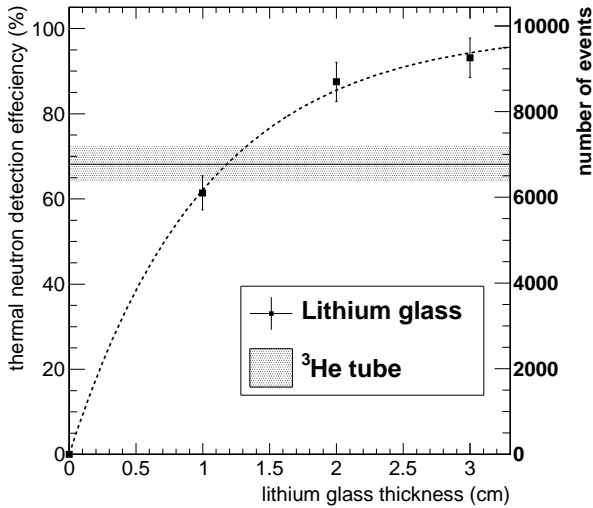}
\caption{\label{fig5} Thermal neutron detection efficiency and
number of events of 1 mm, 2 mm, 3 mm Lithium glass and $^{3}$He tube
in the measurement. using \prettyref{equ:4}. to fit the event number of
different thickness Lithium glass $N_{det}(d)$ to get the total
incident thermal neutron events $N_{0} = 9937\pm698$, the dashed
line is the theoretical relationship between detection efficiency
and detector thickness.}
\end{figure}
\begin{center}
\tabcaption{\label{tab:3}  Number of events and detection efficiency}
\scriptsize
\begin{tabular*}{80mm}{c@{\extracolsep{\fill}}cccc}
\toprule detector & $N_{det}$ & $\epsilon_{det}$ \% & $\epsilon_{det}$(MC) \% \\
\hline
1mm \hphantom{00} & \hphantom{0} $6096 \pm 323$
 & \hphantom{0} $61.4\pm4.0$ & 59.3\\
2mm \hphantom{00} & \hphantom{0} $8694 \pm 477$
 & \hphantom{0} $87.5\pm4.6$ & 85.4\\
3mm \hphantom{00} & \hphantom{0} $9257 \pm 491$
 & \hphantom{0} $93.2\pm4.6$ & 92.9 \\
$^{3}$He tube\hphantom{00} & \hphantom{0} $6777\pm407$ & \hphantom{0} $68.2\pm4.1$ & 70.5\\
\bottomrule
\end{tabular*}
\end{center}
\begin{figure}[H]
\includegraphics[width=7.5cm]{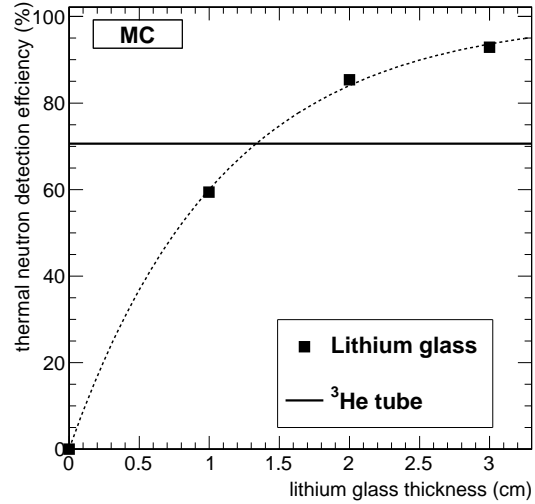}
\figcaption{\label{fig5} Monte Carlo Simulation on thermal neutron
detection efficiency and number of events of 1 mm, 2 mm, 3 mm
Lithium glass and $^{3}$He tube.}
\end{figure}
\section{Conclusion}
We try to find a neutron detector with sufficient detection
efficiency which is designed to enhance electromagnetic calorimeters
in electron/hadron discrimination. Both Lithium glasses and $^{3}$He
tube have been tested with neutron source, and data compared with an
accurate GEANT4 simulation. The 1 mm thickness Lithium glass has a
thermal neutron detection efficiency of about 60\%, almost the same
as the $^{3}$He tube within the error range. The efficiency increase
as the detector thickness, and can reach to 90\% by using the 3 mm
thickness Lithium glass.
Results demonstrate in general good agreement
between data and simulation. The choice of detector for a particular
application is determined by the experiment's requirements with
respect to additional detector characteristics, such as required
size, sensitivity to background radiation such as gamma-rays,
counting rate capability, and position stability as a function of
time. To choose a suitable thermal neutron detector in space
experiments, these factors should all to be considered in further
studies.

\acknowledgments{We would like to thank Dr. Sun Zhi-Jia of China Spallation Neutron Source IHEP team for his suggestions and assisting us in using the $^{3}$He tube.}
\end{multicols}

\vspace{-1mm}
\centerline{\rule{80mm}{0.1pt}}
\vspace{2mm}

\begin{multicols}{2}

\bibliographystyle{apsrev}
\bibliography{bibtex}

\end{multicols}

\clearpage

\end{CJK*}
\end{document}